\def\ts     {\thinspace}
\def\kms    {\ifmmode{{\rm \ts km\ts s}^{-1}}\else{\ts km\ts s$^{-1}$}\fi}
\def\msol     {\ifmmode{{\rm M}_{\odot}}\else{M$_{\odot}$}\fi}

\def\aco  {\ifmmode{^{12}{\rm CO}(J\!=\!1\! \to \!0)}\else{$^{12}{\rm CO}(J\!=\!1\! \to \!0)$}\fi}
\def\bco  {\ifmmode{^{12}{\rm CO}(J\!=\!2\! \to \!1)}\else{$^{12}{\rm CO}(J\!=\!2\! \to \!1)$}\fi}
\def\m    {\ifmmode{\mu {\rm m}}\else{$\mu$m}\fi}
\def\cco  {\ifmmode{^{13}{\rm CO}(J\!=\!1\! \to \!0)}\else{$^{13}{\rm CO}(J\!=\!1\! \to \!0)$}\fi}
\def\dco  {\ifmmode{^{13}{\rm CO}(J\!=\!2\! \to \!1)}\else{$^{13}{\rm CO}(J\!=\!2\! \to \!1)$}\fi}
\def\eco  {\ifmmode{{\rm C}^{18}{\rm O}(J\!=\!1\! \to \!0)}\else{${\rm C}^{18}{\rm O}(J\!=\!1\! \to \!0)$}\fi}
\def\nh   {\ifmmode{N(\hi)}\else{$N$(\hi)}\fi}
\def\hun    {\ifmmode{I_{100}}\else{$I_{100}$}\fi}
\def\sex    {\ifmmode{I_{60}}\else{$I_{60}$}\fi}
\def\hh     {\ifmmode{{\rm H}_2}\else{H$_2$}\fi}
\def\nhh     {\ifmmode{N({\rm H}_2)}\else{$N$(H$_2$)}\fi}
\def\zwco   {\ifmmode{^{12}{\rm CO}}\else{$^{12}{\rm CO}$}\fi}
\def\nzwco   {\ifmmode{N(^{12}{\rm CO})}\else{$N(^{12}{\rm CO})$}\fi}
\def\wzwco   {\ifmmode{W(^{12}{\rm CO})}\else{$W(^{12}{\rm CO})$}\fi}
\def\drco   {\ifmmode{^{13}{\rm CO}}\else{$^{13}{\rm CO}$}\fi}
\def\ndrco   {\ifmmode{N(^{13}{\rm CO})}\else{$N(^{13}{\rm CO})$}\fi}
\def\wdrco   {\ifmmode{W(^{13}{\rm CO})}\else{$W(^{13}{\rm CO})$}\fi}
\def\tex    {\ifmmode{T_{ex}({\rm CO})}\else{$T_{ex}({\rm CO})$}\fi}
\def\ha     {\ifmmode{{\rm H}\alpha}\else{${\rm H}\alpha$}\fi}
\def\amm     {\ifmmode{{\rm NH}_{3}}\else{${\rm NH}_{3}$}\fi}
\def\xco     {\ifmmode{X_{\rm CO}}\else{$X_{\rm CO}$}\fi}
\def\tkin    {\ifmmode{\rm T_{\rm kin}}\else{T$_{\rm kin}$}\fi}
\documentclass{aa}
\usepackage{graphicx}
\begin{document}
   \title{Dense gas in nearby galaxies}

   \subtitle{XV. Hot ammonia in NGC\,253, Maffei\,2 and IC\,342}

   \author{R. Mauersberger\inst{1}
          \and
          C. Henkel\inst{2}
          \and
          A. Wei\ss{}\inst{1,3}
          \and
          A.B. Peck\inst{2,4}
          \and
          Y. Hagiwara\inst{2,5}}

   \offprints{R. Mauersberger (mauers@iram.es)}

   \institute{Instituto de Radioastronom\'{\i}a Milim\'etrica (IRAM), Avda.
Divina
             Pastora 7\,NC, E-18012 Granada, Spain
             \and
             Max-Planck-Institut f\"ur Radioastronomie, Auf dem H\"ugel 71,
             D-53121 Bonn, Germany
             \and
             Radioastronomisches Institut der Universit\"at Bonn,
             Auf dem H\"ugel 69, D-53121 Bonn, Germany
             \and
             Smithsonian Submillimeter Array (SMA), PO Box 824, Hilo, HI
96721, USA
             \and
             ASTRON, Westerbork Observatory, P.O. Box 2, 7990AA Dwingeloo,
The Netherlands}

   \date{Received ; accepted  }

   \abstract{The detection of NH$_3$ inversion lines up to the $(J,K)=(6,6)$ level
   is reported toward the central regions of the nearby galaxies NGC\,253, Maffei\,2,
   and IC\,342. The
   observed lines are up to 406\,K (for $(J,K)=(6,6)$) and 848\,K (for the $(9,9)$
   transition) above the ground state and reveal a warm ($T_{\rm kin}= 100 \ldots 140 $\,K)
   molecular component toward all galaxies studied. The tentatively detected $(J,K)=(9,9)$ line
   is evidence for an even warmer ($>400\,$K) component toward IC\,342. Toward NGC\,253, IC\,342
   and Maffei\,2  the global
   beam averaged NH$_3$ abundances are $1-2\,10^{-8}$, while
   the abundance relative to {\em warm} H$_2$ is around 10$^{-7}$.
   The temperatures and NH$_3$ abundances are similar to values
   found for the Galactic central region. C-shocks
   produced in cloud-cloud collisions can explain kinetic
   temperatures  and
   chemical abundances.  In the central region
   of M\,82, however, the NH$_3$ emitting gas component is comparatively cool
   ($\sim 30\,$K). It must be dense (to provide sufficient NH$_3$ excitation)
   and well shielded from dissociating photons and comprises only a small fraction
   of the molecular gas mass in M\,82.  An important molecular component, which is
   warm and tenuous and characterized by a low ammonia abundance, can be seen mainly in CO.
   Photon dominated regions (PDRs) can explain both the high
   fraction of warm H$_2$ in M\,82 and the observed chemical abundances.
   \keywords{Galaxies: individual: NGC~253, Maffei~2, IC~342, M~82 --
Galaxies: ISM
   -- Galaxies: starburst -- Galaxies: abundances -- Radio lines:
   galaxies}}

   \maketitle
%

\section{Introduction}
Many galaxies, including our own, show large concentrations of
molecular gas in their central few hundred parsecs. Due to the
presence of bars facilitating inflow of gas, this material is
often originating from further out. Reaching the nuclear region,
the gas may trigger bursts of star formation and the formation of
an active nucleus. There have been many studies to determine the
distribution and kinematics of this gas, mainly using the mm-wave
transitions of CO as a tracer of the more elusive H$_2$ (see e.g.
Combes \cite{combes91}, Young \& Scoville \cite{young91}).

In the Galactic disk, there is a well established linear
correlation between the integrated CO intensity, $I_{\rm CO}$, and
H$_2$ (see Sect.\,\ref{x13} for a discussion and references). It
has become evident, however, that the relationship is different in
the central regions of galaxies (e.g. Downes et al.
\cite{Downes92}, Sodroski et al. \cite{sodroski95}, Mauersberger
et al. \cite{mau96a}, \cite{mau96b}, Dahmen et al.
\cite{dahmen98}).

Even more difficult is the determination of local physical
parameters, such as density and kinetic temperature of the gas.
The difficulties compared to similar studies of nearby clouds in
the galactic disk have two main reasons: Firstly, lines of
extragalactic density or temperature tracers, i.e. molecules with
a high dipole moment such as CS, HC$_3$N, or NH$_3$, are broad and
weak; they require high quality baselines and good sensitivity.
Secondly, a low linear resolution forces us to average over
regions of several 100\,pc. Area filling factors are not well
known but are certainly less than unity and it becomes difficult
to disentangle the different gas components which certainly
coexist within one telescope beam.

Nevertheless, there has been progress in determining physical
parameters toward the nuclear regions of nearby galaxies. The most
prominent examples are NGC\,253, M\,82, NGC\,4945 and IC\,342.
These and other galaxies have been mapped with high resolution in
the line emission of high density tracing molecular transitions
(e.g. Downes et al. \cite{Downes92}, Brouillet \& Schilke
\cite{Brouillet93}, Paglione et al. \cite{Paglione95}, Peng et al.
\cite{peng96}). Multilevel studies in CS, HCN, HCO$+$, CH$_3$CN,
CH$_3$CCH, CH$_3$OH, H$_2$CO and HC$_3$N were carried out to
determine H$_2$ densities, $n(\rm H_2)$, toward some of those
galaxies (Mauersberger et al. \cite{mau89}, \cite{mau90},
\cite{mau91}, Walker et al. \cite{walker93}, Jackson et al.
\cite{jackson95}, H\"uttemeister et al. \cite{huette97}, Paglione
et al. \cite{paglione97}, Seaquist et al. \cite{Seaquist98},
Schulz et al. \cite{schulz01}). The level populations of such
molecules depend on their spontaneous deexcitation rates and on
the frequency of collisions with H$_2$, i.e. mostly on the H$_2$
density, and to a lesser degree on the kinetic temperature of the
gas. From these studies of high density tracers we have obtained a
good idea of $N(\rm H_2)$ in components with $n(\rm H_2)$  ranging
between 10$^4$ and several 10$^{5}$\,cm$^{-3}$. In order not to
miss the less dense and more extended gas, studies of CO in its
various transitions across the mm, sub-mm and FIR range were also
carried out (e.g. G\"usten et al. \cite{guesten93}, Mao et al.
\cite{mao2000}, Schulz et al. \cite{schulz01}, Ward et al.
\cite{Ward2001}, Bradford et al. \cite{bradford02}).

One of the less well known parameters has been the gas kinetic
temperature. With area filling factors that are not known, a
thermalized low density tracer like CO cannot be used to determine
$T_{\rm kin}$ in extragalactic sources. Alternative thermometers
for the extragalactic molecular medium are symmetric top molecules
where relative level populations are determined predominantly by
collisions (and to a lesser degree by spontaneous radiation), such
as NH$_3$ or CH$_3$CN. Temperatures in excess of 200\,K have been
obtained for clouds near our Galactic center (Mauersberger et al.
\cite{mau86}, H\"uttemeister et al. \cite{hue93a}, \cite{hue93b},
\cite{hue95}, Rodr\'{\i}guez-Fern\'andez et al.
\cite{rodriguez01}, Ceccarelli et al. \cite{ceccarelli02}). From a
few low energy ammonia lines, kinetic temperatures were determined
toward NGC\,253 (Takano et al. \cite{takano02}), Maffei\,2 (Henkel
et al. \cite{hen00}), IC\,342 (Martin \& Ho \cite{martin86}), and
M\,82 (Wei{\ss} et al. \cite{weiss01a}). However, symmetric top
molecules need H$_2$ densities of order 10$^4$\,cm$^{-3}$ to be
excited. Hence they probe a dense gas component which is important
for star formation but do not trace the temperature of low
density, CO-emitting, molecular gas. For such a low density
molecular component, observations of H$_2$ lines are more
significant. Recently, Rigopoulou et al. (\cite{rigopoulou02})
concluded from observations of rotational H$_2$ lines toward a
number of starburst  and Seyfert galaxies that a significant
fraction (up to 10\% in starburst galaxies; 2---35\% in Seyferts)
of the gas in the centers of such galaxies has temperatures in the
range of 150\,K.

In order to obtain more stringent values for the dense gas and to
identify the origin and excitation of this gas we have measured
higher energy inversion lines up to 405\,K above the ground state
toward the central regions of NGC\,253, Maffei\,2 and up to 848\,K
toward IC\,342. Combined with new CS $J=5-4$ and $^{13}$CO $J=2-1$
data to trace also density and column density, a dense and warm
molecular gas component is revealed in the central regions of
these galaxies.
\begin{table} \caption{Summary of observations}
\label{obssummary}
\begin{tabular}{l l r l }
\hline
Transition & Telescope & Frequency&$\theta_{\rm beam}$ \\
& & GHz& $''$\\ \hline
NH$_3$ (1,1) & MPIfR 100-m &23.694 &40 \\
NH$_3$ (2,2) & MPIfR 100-m &23.723 &40 \\
NH$_3$ (3,3) & MPIfR 100-m &23.870 &40 \\
NH$_3$ (4,4) & MPIfR 100-m &24.139 &39 \\
NH$_3$ (5,5) & MPIfR 100-m &24.533 &39 \\
NH$_3$ (6,6) & MPIfR 100-m &25.056 &38 \\
NH$_3$ (9,9) & MPIfR 100-m &27.478 &34 \\
$^{13}$CO $(2-1)$ & HHT 10-m$^a$ & 220.399& 34\\
CS $(5-4)$   & HHT 10m$^a$ & 244.936&32 \\
\hline
\end{tabular}\\
\begin{footnotesize}
a) in the observed frequency range, the beam efficiency of the HHT
is 0.78
\end{footnotesize}
\end{table}

\begin{figure}
\includegraphics[width=8.5cm]{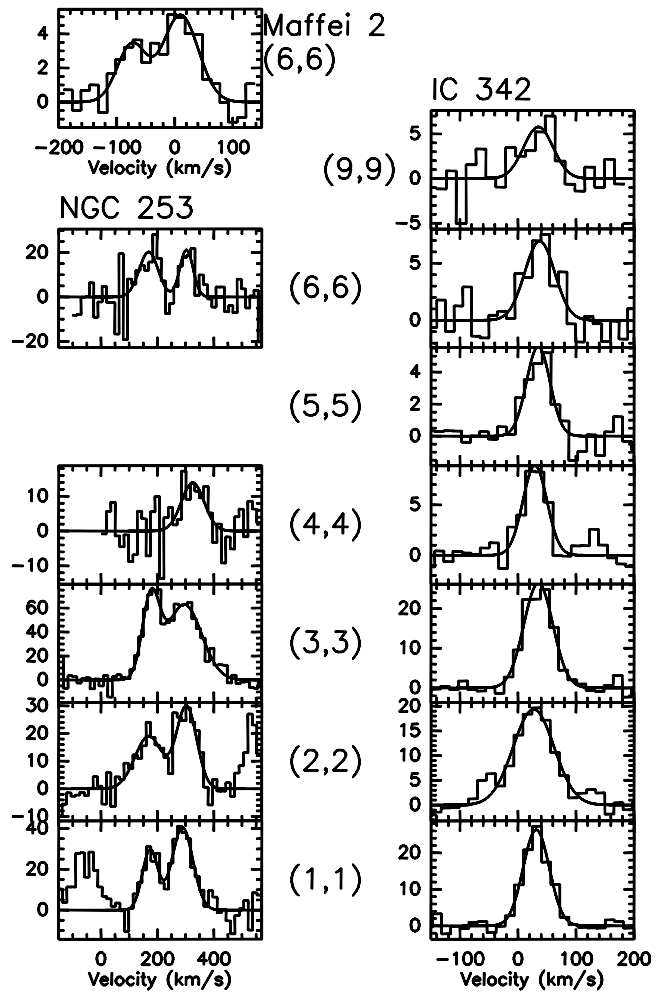}
\caption{Ammonia (NH$_3$) inversion lines (in units of mK, $T_{\rm
mb}$) toward the centers of NGC\,253 ($\alpha_{2000}=0^{\rm
h}47^{\rm m}33\fs2$, $\delta_{2000}=-25^{\rm o}17'16''$),
Maffei\,2 ($\alpha_{2000}=2^{\rm h}41^{\rm m}55\fs2$,
$\delta_{2000}=59^{\rm o}36'11''$) and IC\,342
($\alpha_{2000}=3^{\rm h}46^{\rm m}48\fs6$, $\delta_{2000}=68^{\rm
o}05'46''$), measured at Effelsberg. The spectra have been
smoothed to a velocity resolution of $\sim 15$ km\,s$^{-1}$ (17
km\,s$^{-1}$ for the (9,9) transition). Toward NGC\,253, the
feature to the left of the (1,1) line is part of the (2,2)
profile, and the feature to the right of the (2,2) line is part of
the (1,1) profile.}
\label{spectra}%
\end{figure}

\begin{figure}
\includegraphics[width=9cm]{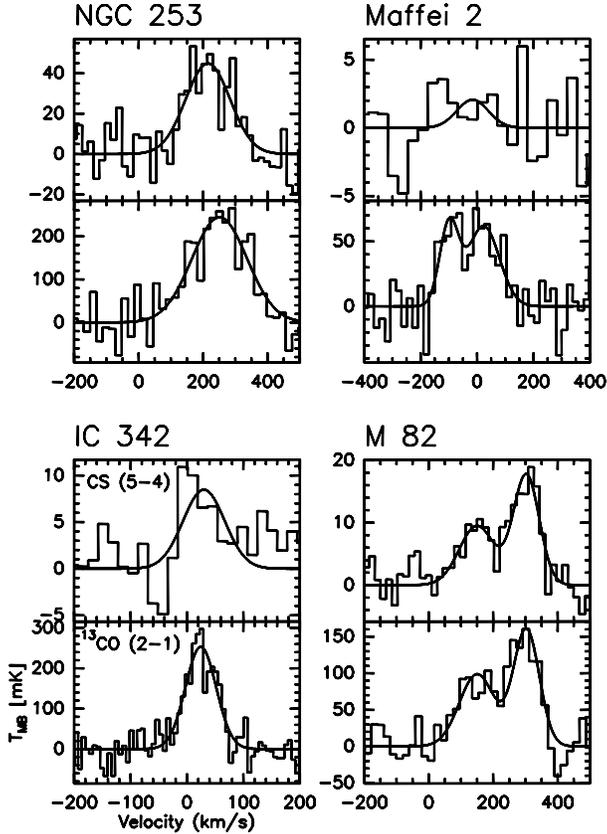}
\caption{CS $J$=5--4 and $^{13}$CO $J$=2--1 profiles measured with
the HHT. For a comparison with NH$_3$ data from Wei{\ss} et al.
(\cite{weiss01a}) we also include profiles measured toward M\,82
($\alpha_{2000}=9^{\rm h}55^{\rm m}52\fs6$, $\delta_{2000}=69^{\rm
o}40'46''$). The velocity resolution has been smoothed to
$17$\,km\,s$^{-1}$. } \label{HHTspectra}
\end{figure}

\begin{table*}
\caption{NH$_3$, CS$(5-4)$ and $^{13}$CO (2-1) results}
\label{results}
\begin{tabular}{l l l l l l l}
\hline Line&$T_{\rm mb}$& $\int T_{\rm mb}{\rm d}v^a$ & $v_{\rm
LSR}^a$&$\Delta v^a$& $N(J,K)^{a,b}$\\
& mK & K\,km\,s$^{-1}$ & km\,s$^{-1}$ & km\,s$^{-1}$ &$10^{12}\rm cm^{-2}$\\
\hline
\multicolumn{6}{l}{\bf NGC\,253}\\
NH$_3$(1,1)&29&2.0(.6) & 174(10) & 64(21) & 26(8) \\
NH$_3($1,1)&41&3.8(.8) &288(8) &87(22) &50(10) \\ \\
NH$_3$(2,2)& 19& 2.3(.7)& 169(15)&114(44)&23(7) \\
NH$_3$(2,2)&29 &2.3(.6) & 303(10)&84(18)&23(6) \\ \\
NH$_3($3,3)&65&5.1(.6) &177(3) &72(6) & 45(5)\\
NH$_3$(3,3)&63& 9.7(.8)& 295(5)&145(13) &84(7)\\ \\
NH$_3$(4,4)$^c$& 14& 1.4(.4)& 325(13)&95(23)&11.2(3.1) \\ \\
NH$_3$(6,6)&20 & 1.7(.5)&{\it 169} & 79(20)& 12.9(3.8)\\
NH$_3$(6,6)&21 & 1.2(.4)&{\it 303} & 55(18)&9.2(3.0)\\ \\
CS$(5-4)$&45&7.7(0.9)&215(10)&162(20)\\
$^{13}$CO$(2-1)$&244&52.4(4.3)&251(8.5)&202(19)\\
 \hline
\multicolumn{6}{l}{\bf Maffei\,2}\\
NH$_3$(6,6)&3.4 &.21(.06) & $-74(8)$&59(17)&1.5 (.4)\\
NH$_3$(6,6)&5.1 &.39(.06) & $10(6)$&72(12)&2.8 (.4)\\
CS$(5-4)$&&$<0.5^d$\\
$^{13}$CO$(2-1)$&6.2&5.2(1.6)&$-99$(10)&79(16)\\
$^{13}$CO$(2-1)$&6.2&8.7(1.9)&23(14)&132(32)\\
 \hline
\multicolumn{6}{l}{\bf IC\,342}\\
NH$_3$(1,1)& 26.3&1.5(.07) & 31.4(1.2)&52.8(2.8) &20.0(1.0)\\
NH$_3$(2,2)& 19.4&1.7(.10) &26.8(2.2) &83.0(6.7) &16.7(1.0)\\
NH$_3$(3,3)&26.4 &1.7(.06) & 35.1(1.1)& 58.8(2.3)&14.9(.5)\\
NH$_3$(4,4)&8.8& .47(.04)&29.1(2.2) &48.9(4.7)  &3.8(.3)\\
NH$_3$(5,5)& 5.6&.29(.03) &34.8(2.6) &47.6(6.2) &2.2(.2)\\
NH$_3$(6,6)&7.0 &.45(.06) & 37.4(3.7)& 60.7(8)&3.2(0.4)\\
NH$_3$(9,9)&5.8&.34(.08)&\it{35}&\it{60}&2.1(.5)\\
CS$(5-4)$&8.5&.78(.17)&\it{30}&86(21)\\
$^{13}$CO$(2-1)$&253&17.2(1.1)&24.5(1.9)&64.1(4.5)\\
\hline
\multicolumn{6}{l}{\bf M\,82}\\
CS$(5-4)$&9&1.3(0.3)&150(11)&129(36)\\
CS$(5-4)$&17&1.8(0.2)&303(6)&94(13)\\
$^{13}$CO$(2-1)$&100&13.5(1.2)&{\it 150}&{\it 129}\\
$^{13}$CO$(2-1)$&150&15.9(1.0)&{\it 303}&{\it 94}\\
 \hline
\end{tabular}\\
\begin{footnotesize}
a) values in italics have been fixed in the Gaussian fits, all errors are 1$\sigma$\\\
b) total beam averaged column density of the observed $(J,K)$
rotational level including both the upper {\em and} lower
inversion state.\\
c) (4,4) data toward NGC\,253 may be affected by pointing problems
(see Sect.\,\ref{nh3results})\\
d) 1$\sigma$ upper limit\\

\end{footnotesize}
\end{table*}
\section{Observations}

\subsection{Effelsberg Observations of NH$_3$ lines}

Between October 2000 and September 2002, the $(J,K)$ = (1,1) to
(6,6) inversion lines of ammonia were observed using the
Effelsberg 100-m telescope\footnote{The 100-m telescope at
Effelsberg is operated by the Max-Planck-Institut f{\"u}r
Radioastronomie (MPIfR) on behalf of the Max-Planck-Gesellschaft}.
The observed line frequencies and the corresponding beam sizes are
summarized in Table\,\ref{obssummary}.

The data were recorded with a dual channel K-band HEMT receiver with $T_{\rm sys}\sim
200\,K$ on a main beam brightness temperature scale ($T_{\rm mb}$). The measurements
were carried out in a dual beam switching mode, with a switching frequency of 1\,Hz
and a beam throw of 2$'$ in azimuth. An `AK\,90' autocorrelator included eight spectrometers
with 512 or 256 channels and bandwidths of 40 or 80\,MHz each, leading to channel spacings of
$\sim$1 or 4\,km\,s$^{-1}$. The eight spectrometers were configured such that the $(J,K)$ =
(1,1) to (4,4), (4,4) and (5,5), or (5,5) and (6,6) lines could be observed simultaneously.
Data with one individual line, either ($J,K$) = (5,5) or (6,6), were also taken. At least
two backends were centered on each ammonia line to sample both linear polarizations
independently. For NGC\,253, only one of the two receiver channels provided useful data.

The (9,9) line at 27.478\,GHz was observed in December 1995 using
a single channel 1\,cm HEMT receiver equipped with a 1024 channel
autocorrelator. System temperatures were near 500\,K on a $T_{\rm
mb}$ scale. The beam size was $34''$. The (9,9) line was measured
in a position switching mode with a bandwidth of 25\,MHz and a
channel spacing of $\sim$0.27\,km\,s$^{-1}$.

Flux calibration was obtained by regularly observing both the continuum and the ammonia
lines of W3(OH) (for fluxes see Mauersberger et al. \cite{mau88}, Ott et al. \cite{ott94}).
Pointing was checked every hour. W3(OH) was the pointing source for IC\,342 and Maffei\,2;
PKS\,$0023-26$ was used for NGC\,253. The pointing accuracy was found to be stable within
5$''$--10$''$.
%

A linear baseline was removed from each individual spectrum and
spectra were added with relative weights depending on the noise
level outside the velocity range of the respective line. We
estimate the flux calibration of the final reduced spectra to be
accurate within $\pm$15\%.

\subsection{CS and $^{13}$CO observations with the Heinrich-Hertz-Telescope}

The $J$=5--4 transition of $^{12}$C$^{32}$S (hereafter CS) and the
$J$=2--1 transition of $^{13}$C$^{16}$O (hereafter $^{13}$CO) were
observed in November 2001 toward NGC\,253, Maffei\,2, IC\,342, and
M\,82 using the 10-m Heinrich-Hertz-Telescope\footnote{The HHT is
operated by the Submillimeter Telescope Observatory on behalf of
Steward Observatory and the MPIfR.} (Baars \& Martin
\cite{Baars96}). At the line frequencies of 244.9536\,GHz (CS) and
220.3987\,GHz ($^{13}$CO), the beam width was $\sim$32$''$. The
zenith optical depth at 225\,GHz as determined by a tipping
radiometer was typically 0.2---0.4. A single channel SIS receiver
with a total bandwidth of $\sim$0.8\,GHz tuned in double sideband
mode was used for the observations. The spectrometer was a 2048
channel AOS with a channel separation of 0.48\,MHz
($\sim$0.6\,km\,s$^{-1}$). The beam efficiency of the HHT is 0.78
in the observed frequency range. System temperatures were
600---1000\,K on a $T_{\rm mb}$ scale. The calibration procedure
was described by Mauersberger et al. (\cite{mau99}). The observing
mode was symmetric wobbler switching with a throw of $\pm 200''$
in azimuth. The pointing was checked every 3---5 hours and was
accurate to $<$5$''$. Errors in the calibration procedure,
atmospheric fluctuations and deviations from a sideband gain ratio
of unity should result in an uncertainty of $\pm 20\%$ in the
intensity scale. After co-adding the spectra, we subtracted first
order baselines and smoothed to a velocity resolution of $\sim 17$
km\,s$^{-1}$.

\section{Results}

\subsection{NH$_3$}
\label{nh3results}

The measured spectra are displayed in Fig.\,\ref{spectra}, and the
results of Gaussian fits to the profiles, namely the velocity
integrated main-beam brightness temperatures $\int T_{\rm mb}{\rm
d }v$, the velocities relative to the local standard of rest
($v_{\rm LSR}$) and the full-width to half power linewidths
($\Delta v_{1/2}$) are summarized in Table\,\ref{results}. For the
(9,9) line toward IC\,342 we have fixed the central velocity and
linewidth in order to reduce the nominal error for the integrated
intensity. We have clearly detected the (1,1) \ldots (6,6)
transitions and tentatively also the (9,9) transition toward
IC\,342, the (1,1) \ldots (4,4) and (6,6) lines toward NGC\,253
and the (6,6) line toward Maffei\,2. These are the first detection
of the (5,5) transition, the first definite detections of the
(6,6) transition and the first tentative detection of an NH$_3$
(9,9) line outside the Galaxy.

Toward NGC\,253, we find two velocity components in the (1,1),
(2,2), (3,3) and (6,6) lines at $\sim$175 and 290\,km\,s$^{-1}$,
each with a linewidth of 80--100\,km\,s$^{-1}$, which presumably
arise from two intensity peaks also seen in higher resolution maps
(Mauersberger et al. \cite{mau96a}, Harrison et al.
\cite{harrison99}). These velocity components cannot be
distinguished in our (4,4) profile. As line width and central
velocity indicate, this line only shows the 300 km\,s$^{-1}$
component. Therefore we cannot exclude an exceptionally large
pointing error favoring the higher velocity peak. From CO mapping
(e.g. Mauersberger et al. \cite{mau96a}, Dumke et al.
\cite{dumke2001}) the observed line profile can be reproduced if
the telescope was mispointed by $\sim 10''$ toward the SW.
Therefore we will not account for the detailed shape of the (4,4)
line in the further interpretation of our results but will take
the presence of the (4,4) line as just another indication for the
presence of warm molecular gas.

For the lower lying transitions of ammonia and other molecules toward Maffei\,2, there are
two velocity components, at about --80 and +6\,km (e.g. Henkel et al. \cite{hen00}), each
with a line width of $\sim 50$ km\,s$^{-1}$. Our (6,6) spectrum is consistent with these results
(see Fig.\,\ref{spectra} and Table 1). A single velocity component could fit all data from IC\,342.

\subsection{CS $(5-4)$ and $^{13}$CO $(2-1)$}

We have detected CS $(5-4)$ emission toward NGC\,253, IC\,342, and
M\,82. Toward Maffei\,2, we obtain an upper limit. $^{13}$CO
emission was observed toward all four galaxies. The profiles are
shown in Fig.\,\ref{HHTspectra} and results from Gaussian fits are
presented in Table\,\ref{results}.
\subsection{Source structure, beam size and line intensities}
\label{structure} To constrain molecular abundances and physical
conditions from observed CS line intensities, it is crucial that
the transitions are observed with comparable beam sizes or that
appropriate corrections are applied accounting for beam size and
source morphology. Not included in this discussion is Maffei\,2
for which not enough CS data exist. For IC\,342 we have assumed
that the observed CS and NH$_3$ emission arises from the same
volume as the $^{13}$CO $J$=1--0 emission that was observed with
high spatial resolution by Meier et al. (\cite{meier00}).
Smoothing the $^{13}$CO $J$=1--0 data cube to different beam sizes
thus enables us to investigate the dependence of observed line
intensity on the beam size. The same procedure was applied to the
distribution of the molecular gas in M\,82 using the high angular
resolution C$^{18}$O $J$=1--0 map of Weiss et al.
(\cite{weiss01b}) and to NGC\,253, using the distribution of CS
$J$=2--1 (Peng et al. \cite{peng96}). The CS line intensities
expected at $32''$ resolution (the beam size of our CS $J$=5--4
data, see Sect.\,2.2) are summarized in
Table\,\ref{cs-intensities}. Given the uncertainties of the latter
approximation and accounting for possible differences between the
C$^{18}$O or $^{13}$CO versus the CS distributions in M\,82 and
IC\,342, we estimate an error of $\pm 30\%$ for the intensity of
individual CS lines convolved to a beam size of 32$''$.

\begin{table}
\caption{CS peak line intensities in main beam brightness
temperature (mK) at $32''$ resolution, the HHT CS $J$ =5--4 beam
size. Estimated errors are $\pm 30\%$ (see
Sect.\,\ref{structure}).} \label{cs-intensities}
\begin{tabular}{l l l l}
\hline Transition  & NGC\,253  & IC\,342         & M\,82 (SW)\\
\hline
CS\,$J=2-1$        &160$^a$    & 80$^a$          & 92$^d$ \\
CS\,$J=3-2$        &77$^a$     & 45$^{a,e}$      & 34$^a$ \\
CS\,$J=5-4$        &50$^{a}$,\,45$^{b}$ & 12$^{a}$,\,$8.5^{b}$  &
13$^{b,e}$
\\
C$^{34}$S\,$J=2-1$ &23$^a$     & 12$^{c,f}$      & 14$^a$ \\
C$^{34}$S\,$J=3-2$ &--         &  8$^{c,f}$      & 4$^a$ \\ \hline
\end{tabular}\\
\begin{footnotesize}
a) original data from Mauersberger \& Henkel (\cite{mau89})\\
b) this paper\\
c) original data from Mauersberger et al. (\cite{mau95})\\
d) original data from Baan et al. (\cite{Baan89})\\
e) data measured with $0'',+10''$ offset to other lines\\
f) data measured with $0'',-10''$ offset to other lines\\
\end{footnotesize}
\end{table}

\section{Discussion}

In the following, we discuss the nuclear regions of all four
galaxies observed in the $^{13}$CO $J$=2--1 and CS $J$=5--4
transitions, i.e. NGC\,253, Maffei 2, IC\,342, and M\,82 (see
Fig.\,\ref{HHTspectra}). NH$_3$ data from M\,82 will be taken from
Wei{\ss} et al. (\cite{weiss01a}). The NH$_3$ observations
(Fig.\,\ref{spectra}, Wei\ss{} et al. \cite{weiss01a}) will be
used to derive kinetic temperatures in regions dense enough ($\ga
10^4$\,cm$^{-3}$) to excite the NH$_3$ inversion lines. The
$^{13}$CO observations are needed to estimate H$_2$ column
densities and the CS data will be used to determine $n$(H$_2$)
densities, CS column densities, and CS abundances relative to
H$_2$.
\subsection{NH$_3$ column densities and rotational
temperatures} \label{ammonia}

\begin{figure*}
\includegraphics[angle=-90, width=14cm]{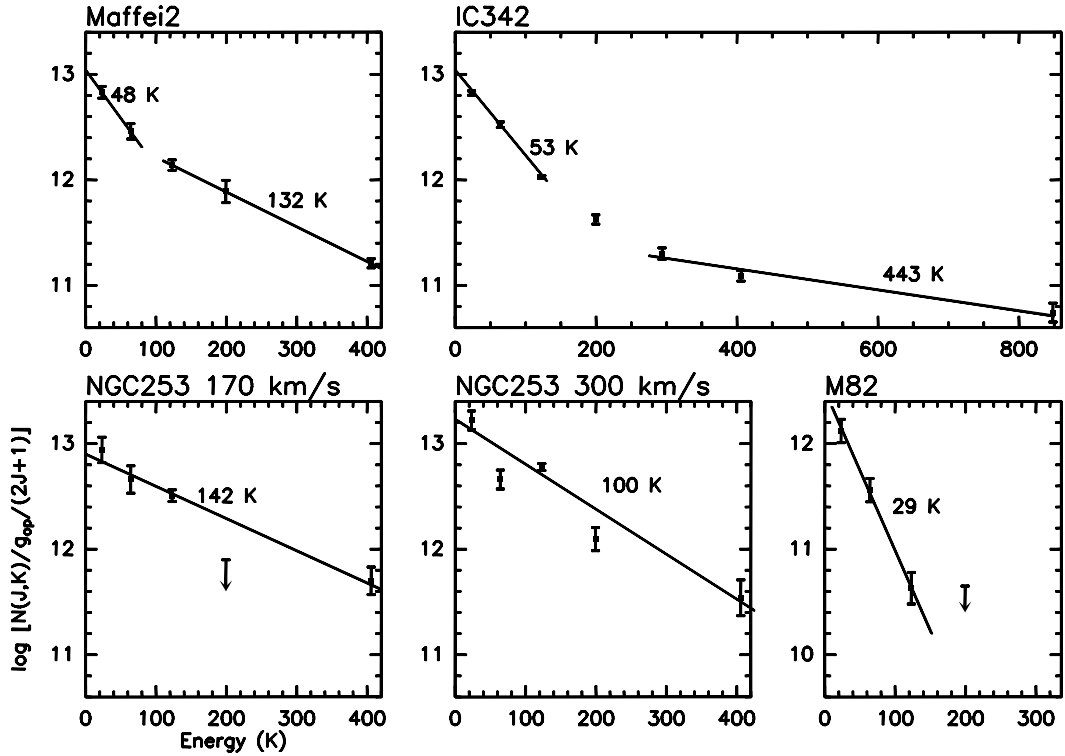}
\caption{A Boltzmann plot for the metastable NH$_3$ levels (i.e.
with $J=K$) observed toward Maffei\,2 (data for $(J,K)<(6,6)$ are
from Henkel et al. \cite{hen00}), for IC\,342 and for the two
velocity components of NGC\,253 at 170 and 300 km\,s$^{-1}$. We
also show the M\,82 data from Wei\ss{} et al. (\cite{weiss01a})
correcting an error in the position of the (4,4) limit in their
paper. The x-axis denotes the energy level of the corresponding
rotational transition in K, the y-axis denotes $\log
(N(J,K))/g_{\rm op}(2J+1) $ (see Sect.\,\ref{ammonia}). For a
better comparison of the resulting temperatures, the scales in all
frames and axes are the same. Linear regressions to the data and
the resulting values for $T_{\rm rot}$ are shown. \label{boltz} }
\end{figure*}

In the optically thin case, the beam averaged column density of a
rotational state of ammonia (i.e. upper plus lower inversion
level) is related to the velocity integrated main-beam brightness
temperature by
\begin{equation}
N(J,K)=\frac{1.55\,10^{14}\rm cm^{-2}}{\nu}\frac{J(J+1)}{K^2}\int
T_{\rm mb} {\rm d}v,
\end{equation}
with $\nu$ being the line frequency in GHz, and $\int T_{\rm mb}
{\rm d}v$ being the integrated intensity in $\rm K\,km\,s^{-1}$.
Here we assumed that the excitation temperature of the inversion
lines is $>1\rm K$, the energy difference of the two states of a
given $(J,K)$ level. The calculated total column densities and
their errors are displayed in Table \ref{results}. Because most
lines were measured simultaneously, the main contribution to the
given uncertainties arises from the errors in the Gaussian fits.

If one plots $\log_{10}(N(J,K)/(g_{\rm op}(2J+1)))$, where
$g_{op}=2$ for ortho-ammonia ($K=3, 6, 9, \ldots$) and  1 for
para-ammonia ($K=1,2,4,5,7 \ldots$), against the energy of the
involved $(J,K)$ rotational state, the slope of a linear fit to
two or more rotational levels, $m$, is related to a rotational
temperature ($T_{\rm rot}$) by $T_{\rm rot}=-0.434/m$ (e.g.
Mauersberger et al. \cite{mau86}). Such rotational diagrams
(Boltzmann plots) are displayed in Fig.\,\ref{boltz} for the
observed galaxies. For NGC\,253, a Boltzmann plot is given for
each of the two observed velocity components. Column densities for
Maffei\,2 have been calculated from the data in Henkel et al.
(\cite{hen00}).


Fig.\,\ref{boltz} shows the first extragalactic Boltzmann plots
from ammonia with fits extending beyond the (4,4) line (energy
level: 199\,K) to energies of 405\,K and 848\,K.
Table\,\ref{summary} summarizes data for the central regions of
the galaxies discussed in more detail below. Model calculations
(e.g. Walmsley \& Ungerechts \cite{walmsley83}) indicate that due
to collisional deexcitation of metastable levels (i.e. with $J=K$)
via lower lying metastable levels the rotational temperature is
only a lower limit to the gas kinetic temperature. This effect is
less pronounced for the higher metastable ammonia lines, which
therefore should be  a more reliable indicator of kinetic
temperatures.
\subsubsection{Comments on individual sources}
\paragraph{NGC\,253.} For NGC\,253, the $(4,4)$ line has not been included
in the fit (see Sect.\,\ref{nh3results}). We show fits for two
velocity components, each corresponding to one of the twin peaks
separated symmetrically by about 10$''$ from the center of
NGC\,253 (Mauersberger et al. \cite{mau96a}). The data can be
fitted by a single rotational temperature of 142($\pm 14$)K  for
the 170\,km\,s$^{-1}$ component and 100($\pm 3$)\,K for the
300\,km\,s$^{-1}$ component. These values are much higher than the
rotational temperature of 17--50\,K derived from measurements of
the $(1,1)$ to $(3,3)$ lines and a non-detection of the (4,4) line
by Takano et al. (\cite{takano02}). Ortho to para ammonia
abundance ratios of up to 6 have been claimed by Takano et al.
(\cite{takano02}) for some velocity components toward NGC\,253.
Such a high ratio would indicate that NH$_3$ was formed in a very
cold ($<10$\,K) environment. Although the (3,3) line seems to be
stronger than expected from the intensities of the (2,2) and (4,4)
lines, the (6,6) line (which also belongs to the ortho species)
has a normal intensity. We therefore cannot confirm an anomalous
ortho-to-para ratio in NGC\,253. Unlike in Maffei\,2 and IC\,342
(see below), in NGC\,253 we cannot identify an important amount of
cool ($<50$\,K) in ammonia. In Sect.\ref{originand heating} we
argue that this is due to an elevated relative abundance of NH$_3$
in the warm component.

If we assume a thermalization at those temperatures and further
assume that the non-metastable ammonia levels (i.e. those with
$J>K$) are not populated (a reasonable assumption because a
significant population would require $n(\rm H_2)>10^6\,\rm
cm^{-3}$, e.g. Mauersberger et al. \cite{mau85}), the total
ammonia column densities obtained are 1.6\,10$^{14}$\,cm$^{-2}$
for the 170\,km\,s$^{-1}$ and 2.1\,10$^{14}$\,cm$^{-2}$ for the
300\,km\,s$^{-1}$ component.
\paragraph{Maffei\,2.}  As toward NGC\,253, two velocity components are
detected in Maffei 2. While Henkel et al. (\cite{hen00}) suggested
that the --80\,km\,s$^{-1}$ component shows higher excitation than
the +6\,km\,s$^{-1}$ feature, our (6,6) line spectrum
(Fig.\,\ref{spectra}) indicates equal excitation within the limits
of observational accuracy. A single rotational temperature does
not yield a good fit to the data. The (1,1) and (2,2) lines can be
fitted by  $T_{\rm rot}=48(\pm 15)\,$K and the (3,3) to (6,6)
lines by $T_{\rm rot}=132(\pm 12)$\,K. In a previous study, Takano
et al. (\cite{takano00}) derived from a stronger than expected
(3,3) line of ortho-ammonia an anomalously high ortho/para ratio
of 2.6 for ammonia, which would be an indication of NH$_3$
formation in a cold (13\,K) environment. Our higher
signal-to-noise (4,4) and (6,6) data do not support this finding:
the ortho/para ratio toward Maffei\,2 is normal. Assuming that the
(0,0) rotational level is connected to the (1,1) and (2,2) levels
by a rotational temperature of 48\,K, the total ammonia column
density in all metastable levels is
8.6(1.3)\,10$^{13}$\,cm$^{-2}$.
\paragraph{IC\,342.} There is no single
rotational temperature for all the observed lines. The $(1,1)
\ldots (3,3)$ transitions can be represented by $T_{\rm
rot}=53(\pm 1)\,\rm K$ and the $(5,5) \ldots (9,9)$ lines by
$T_{\rm rot}=443(\pm 130)\,\rm K$. If we had not included the
(9,9) data, the temperatures obtained would have been similar to
those obtained for NGC\,253. Assuming that the (0,0) rotational
level is connected to the $(1,1) \ldots (3,3)$ levels by a
rotational temperature of 53\,K, the total ammonia column density
in all metastable levels is 7.5(0.5)\,10$^{13}$\,cm$^{-2}$. About
1/3 of this total column belongs to a hot component of several
100\,K.
\paragraph{M\,82.} Although we are adding no new ammonia data to
those observed by Wei\ss{} et al. (\cite{weiss01a}), we notice
that in their Boltzmann plot the limit for the (4,4) line was
erroneously plotted too low. While the (1,1) \ldots (3,3) lines
can be described by a common temperature of $29(\pm 1)\,$K, the
(4,4) line's upper limit cannot exclude that the higher ammonia
levels are described by components with high rotational
temperature, like in other galaxies observed.
\subsection{Determining $N(\rm H_2)$ from $^{13}$CO \label{x13}}
The integrated intensity of $^{12}$CO $1-0$ has been empirically
related to the H$_2$ column density for Galactic disk molecular
clouds (e.g. Dickman \cite{Dickman75}, Sanders et al.
\cite{sanders85}, Bloemen et al. \cite{Bloemen86}). The
theoretical explanation of such a universal $X$-factor requires
three conditions to be fulfilled (Dickman et al. \cite{Dickman86},
Mauersberger \& Henkel \cite{mau93}), namely that (a) there are
many unresolved clouds with optically thick CO emission within a
beam (CO counts clouds), (b) that these clouds are in virial
equilibrium and (c) that the CO rotational levels are populated
via collisions with H$_2$ molecules. For molecular clouds near
galactic centers (including our own), conditions a) and c) are
likely fulfilled. It is, however, by no means clear whether
molecular clouds close to the central regions of galaxies are
virialized (except perhaps the very densest regions). Tidal forces
from the nuclear regions or high stellar densities may easily
rival the proper gravitational forces of a molecular cloud near a
galactic center (e.g. G\"usten \cite{guesten89}, Mauersberger et
al. \cite{mau96a}). For such tidally disturbed clouds, the
theoretical explanation of the $X$ factor, which works so well for
molecular clouds in the Galactic disk, would not be valid anymore.
This is even more true for any molecular intercloud medium (e.g.
Downes et al. \cite{Downes92}, Jog \& Das \cite{jog93}, Solomon et
al. \cite{solomon97}, Downes \& Solomon \cite{downes98}), which is
certainly not virialized. An increasing number of examples reveal
that the same $X$ factor from the Galactic disk cannot be used
near the centers of galaxies (Galactic Center: Sodroski et al.
\cite{sodroski95}, Dahmen et al. \cite{dahmen98}, NGC\,253:
Mauersberger et al. \cite{mau96a}, NGC\,4945: Mauersberger et al.
\cite{mau96b}, M\,82: Mao et al. \cite{mao2000}, Wei\ss{} et al.
\cite{weiss01b}, ultraluminous galaxies: Solomon et al.
\cite{solomon97}).

Relating optically thin transitions of CO to the H$_2$ column
density requires far fewer assumptions than for optically thick
$^{12}$CO transitions. The mm-transitions of $^{12}$C$^{18}$O
(hereafter C$^{18}$O) are in most conceivable cases of
astrophysical interest optically thin (e.g. Mauersberger et al.
\cite{mau92}); the transitions of $^{13}$CO should be in most
cases optically thin or only moderately optically thick. Because
of the high chemical stability of CO and its isotopes their
relative chemical abundance should vary less than that of other
molecules in the interstellar medium. Model calculations
(Mauersberger et al. \cite{mau92}) show that for a large range of
temperatures and densities the intensity $I=\int T_{\rm mb}{\rm
d}v$ (in K\,km\,s$^{-1}$) of an optically thin isotopic CO
($^{i}$C$^j$O) $J=2-1$ line is related to the H$_2$ column density
by
\begin{equation}
\label{equation} N({\rm H_2})=5.3\,10^{14} {\rm
cm^{-2}}(\frac{[^i{\rm C}^j{\rm O}]}{[\rm H_2]})^{-1}I(^i{\rm
C}^j{\rm O}).
\end{equation}
Any increase in $T_{\rm ex}$ of the corresponding levels due to
higher excitation (i.e. higher $n(\rm H_2)$, $T_{\rm kin}$) is
approximately compensated by a loss of level population of the
corresponding states to higher states in the $J$ ladder; this is
why Eq.\,\ref{equation} holds even if the $J=(2-1)$ line is
subthermally excited. An unknown quantity in Eq.\,\ref{equation}
is the relative abundance of the CO isotopomers studied. For
Galactic disk molecular clouds, a  CO abundance of 8\,10$^{-5}$ is
frequently assumed (Frerking et al. \cite{frerking82}). Toward the
centers of spiral galaxies, metallicities are typically higher by
a factor of two than in the local disk (see Vila-Costas \& Edmunds
\cite{vila92}). On the other hand in the centers of galaxies the
amounts of carbon in the form of C\,\textsc{I} and CO may be
comparable (Gerin \& Phillips \cite{gerin2000}, Israel \& Baas
\cite{Israel02}). It is therefore plausible to expect for the
galaxies studied here a global CO/H$_2$ abundance similar to the
Galactic disk value, with an uncertainty which could be a factor
of two in either direction.

Model calculations suggest that CO isotopomers are not very
greatly affected by selective dissociation (Chu \& Watson
\cite{chu83}, Glassgold et al. \cite{glassgold85}). A
fractionation of $^{13}$CO via the reaction
\begin{equation}
^{12}\rm CO + ^{13}CO^+ \rightarrow ^{13}CO + ^{12}C +\Delta \rm
E_{\rm 35 K} \end{equation} (Watson \cite{watson76}) becomes
significant only for kinetic temperatures smaller than 35\,K (e.g.
Langer et al. \cite{langer84}). As long as the kinetic
temperatures of the CO emitting gas are higher than this it is
safe to assume that fractionation of $^{13}$CO introduces an error
of the H$_2$ column density which is negligible in comparison to
the uncertainty of the CO/H$_2$ abundance.

The $^{12}$C/$^{13}$C and $^{16}$O/$^{18}$O ratios have been
determined to be 50 and 200 (Henkel \& Mauersberger
\cite{henkel93}, Wei\ss{} et al. \cite{weiss01b}) in the galaxies
studied. Eq.\,(1) transforms into $N(\rm H_2)$ =
1.3\,10$^{21}\,{\rm cm^{-2}}$\,$I({\rm C}^{18}{\rm O})$ and $N(\rm
H_2)$ = 3.3\,10$^{20}\,{\rm cm^{-2}}$\,$I(^{13}{\rm C}{\rm O})$.
We have used the latter equation together with our measurements of
the $^{13}$CO $J=2-1$ line to determine the beam averaged H$_2$
column densities summarized in Table\,\ref{summary}. If the
$^{13}$CO lines are slightly optically thick, the $N(\rm H_2)$
values may be higher (see below). While C$^{18}$O lines could be
an even more accurate tracer of $N(\rm H_2)$, such lines are
substantially weaker and are difficult to detect with a proper
signal-to-noise ratio.

A comparison between the derived $\rm H_2$ column densities and
$^{12}$CO $J=1-0$ measurements (accounting for different spatial
resolutions, see Sect\,\ref{structure}) towards NGC\,253 (Houghton
et al. \cite{hou97}, Sorai et al. \cite{sor00}) and IC\,342
(Crosthwaite et al. \cite{cro01}, Meier \& Turner \cite{meier01})
shows, that a conversion factor $X_{\rm CO}$ which is a factor of
$3-5$ smaller than the Galactic value of $2.3\,10^{20}\,{\rm
cm^{-2} \,(K \kms)^{-1}}$ (Strong et al. \cite{strong88}) is
appropriate for starburst nuclei. This is in line with recent
determinations of $X_{\rm CO}$ in M\,82 (Wei\ss{} et al.
\cite{weiss01b}).

\subsection{$n(\rm H_2)$ and CS abundance}
\begin{table*}
\caption{Physical parameters in the central regions ($\sim 35''$)
of nearby galaxies} \label{summary}
\begin{tabular}{l l l l l l l l l l}
\hline Source &$D$ & 35$''$ equiv. &$N({\rm H_2})^a$&$\log n(\rm
H_2)$ &$T_{\rm rot}$
&$N({\rm NH_3})$&$N({\rm CS})$&log($X{(\rm NH_3)}$)&log$(X{(\rm CS)})$\\
&&beam size&&&\\
&Mpc& pc&10$^{21}$\,cm$^{-2}$&cm$^{-3}$ &K& \multicolumn{2}{c}{10$^{13}$\rm cm$^{-2}$}& &\\
\hline
NGC\,253&2.5$^{b}$ &420&17.4 &$4.2\pm1.1$&100\ldots 142&37(7)&$2-20$&$-7.7(0.3)$&$-8.4\pm0.5$\\
Maffei\,2 &2.5$^c$&420&4.6& &48, 132&8.6(1.3)& &$-7.7(0.3)$&\\
IC\,342 &1.8$^d$&300 &5.7&$3.9\pm0.3$& 53, 443 &7.5(.5)&$1-6$&$-7.9(0.3)$&$-8.4\pm0.5$\\
M\,82& 3.9$^e$&660&9.7&$4.3\pm1.0$ & 29$^f$ & 1.0(.3)$^f$ &$1-60$
&$-9.0(0.3)$&$-7.9\pm0.8$\\ \hline
\end{tabular}
\\
\footnotesize{ a) from our $^{13}$CO $J=2-1$ data; the uncertainty
is a factor of 2 in both directions. b) Mauersberger et al.
\cite{mau96a}; c) Karachentsev et al. \cite{kara97}; d) McCall
\cite{mccall89}; e) Sakai \& Madore \cite{sakai99}; f) Wei\ss{} et
al. \cite{weiss01b}}
\end{table*}
We now use the intensity of the CS $(5-4)$ lines given in
Table\,\ref{cs-intensities} to complement the multilevel studies
in Mauersberger \& Henkel (\cite{mau89}). The levels corresponding
to this transition need an H$_2$ density  $\sim 10^4 \rm
\,cm^{-3}$ to be excited. We used the large velocity gradient
(LVG) code described in Mauersberger \& Henkel (\cite{mau89}) to
investigate the \hh{} density, the CS column density and relative
abundance ($n(\hh)$, $N(\rm CS)$ and $X$(CS)=$N$(CS)/$N$(H$_2$)).
Note that the values given below for $X$(CS) are beam averaged
values for the CS emitting dense gas component (the LVG code gives
the source averaged $N$(CS) which can be multiplied with the beam
filling factor, given by a comparison of the code's results with
the observed intensities, to obtain the beam averaged $N$(CS)).
Any lack of CS emission in the intercloud component due to a low
abundance or insufficient excitation increases the local abundance
in the CS emitting component. LVG models were calculated for
$\log(n(\hh))\,=2.0$ to 6.0 in steps of 0.2 and $\log(X({\rm
CS})/{\rm grad}(v))=-11.7$ to $-6.7$ in steps of 0.2. The relative
abundance [CS]/[C$^{34}$S] was fixed to 25, the Galactic
$^{32}$S/$^{34}$S abundance ratio (Chin et al. \cite{chin96}).
Since the kinetic temperature of the gas is high (see the results
in Sect. \ref{ammonia}) and the excitation of CS does only weakly
depend on $T_{\rm kin}$ for temperatures above 30\,K (see e.g.
Mauersberger \& Henkel \cite{mau89}) we fixed $T_{\rm kin}$ to
60\,K. Solutions were selected by comparing all beam corrected CS
and C$^{34}$S line intensity ratios derived from
Table\,\ref{cs-intensities} to the corresponding LVG line
intensity ratios using a $\chi^2$-test. Only those solutions were
permitted where the beam averaged \hh\ column densities were not
below the values derived from our $^{13}$CO $J=2-1$ measurements
(NGC\,253 and IC\,342).

For Maffei\,2, our non-detection of the $J$=5--4 line and the
intensity of the $J$=3--2 CS line detected by Mauersberger et al.
(\cite{mau89a}) do not permit a reliable determination of
$n$(H$_2$). For IC\,342 all observed line intensities can be
fitted simultaneously by the LVG models. We derive a moderate
$n$(H$_2$) density of $10^{3.9\pm0.3}\rm cm^{-3}$. The beam
averaged CS column density is between $1\,10^{13}$\,cm$^{-2}$ and
$6\,10^{13}$\,cm$^{-2}$. Using the H$_2$ column density given in
Table\,\ref{summary} this corresponds to a beam averaged relative
CS abundance of $10^{-8.4\pm 0.5}$. From an HCN multilevel study
Paglione et al (\cite{paglione97}) estimated elevated kinetic
temperatures and densities toward IC\,342 which are consistent
with our models. Using more transitions of HCN, Schulz et al.
(\cite{schulz01}) point out that 5\% of the dense gas might have
densities of 10$^6$\,cm$^{-3}$ and might be as cool as 30\,K.

For NGC\,253 and M\,82 not all observed line intensities can be
fitted by the LVG models simultaneously. As discussed in
Mauersberger \& Henkel (\cite{mau89}) for NGC\,253 this indicates,
that the lower transitions of CS arise from a lower density
component containing dense concentrations traced by the $5-4$
transition. This explanation is supported by the LVG solutions we
obtain when we do not include either the CS $J=2-1$ or CS $J=3-2$
transition: for NGC\,253 we obtain $n(\hh)=10^{4.2\pm0.7}\,\rm
cm^{-3}$ (using CS $J=2-1$, CS $J=5-4$ and C$^{34}$S $J=2-1$) and
$n(\hh)=10^{4.5\pm0.8}\,\rm cm^{-3}$ (using CS $J=3-2$, CS $J=5-4$
and C$^{34}$S $J=2-1$). Including only the low $J$ transitions
yields $n(\hh))=10^{3.7\pm0.6}\,\rm cm^{-3}$. The beam averaged CS
column density is between $2\,10^{13}\,{\rm cm^{-2}}$ and
$2\,10^{14}\,{\rm cm^{-2}}$. The beam averaged CS abundance
therefore is $10^{-8.4\pm0.5}$. Our density estimates are
consistent with multilevel studies of HCN (Paglione et al.
\cite{paglione97}), HC$_3$N (Mauersberger et al. \cite{mau90}),
CH$_3$CN and CH$_3$CCH (Mauersberger et al. \cite{mau91}).

The corresponding results for M\,82 are:
$n(\hh))=10^{3.9\pm0.7}\,\rm cm^{-3}$ (using CS $J=2-1$, CS
$J=5-4$, C$^{34}$S $J=2-1$, and C$^{34}$S $J=3-2$) and
$n(\hh))=10^{4.2\pm0.3}$ (using CS $J=3-2$, CS $J=5-4$ and
C$^{34}$S, $J=2-1$ and C$^{34}$S $J=3-2$). The density is not well
constrained using the low $J$ lines only. LVG solutions then cover
a density range from $n(\hh)=10^{3.3}$ to $10^{5.5}\,\rm cm^{-3}$
and low \hh\ densities are associated with high `local' CS
abundances and vice versa. Interestingly, solutions are only
permitted for $T_{\rm kin}<25$\,K. Beam averaged CS column
densities including the CS $J=5-4$ line are in the range between
$1\,10^{13}\,{\rm cm^{-2}}$ and $6\,10^{14}\,{\rm cm^{-2}}$. The
beam averaged CS abundance is $10^{-7.9\pm0.8}$. The LVG results
are summarized in Table\,\ref{summary}. Our density estimates are
consistent with those from HCN (Paglione et al. \cite{paglione97})
and CH$_3$CCH (Mauersberger et al. \cite{mau91}).

\subsection{Origin and heating of the warm gas}
\label{originand heating} In the complex environment of the
central few 100\,pc of an active galaxy there are certainly
several coexisting gas components. Transitions from molecules with
a high dipole moment such as ammonia and CS need a density $\sim
10^4\,\rm cm^{-3}$ to be excited, in contrast to  CO lines which
need a few 100\,$\rm cm^{-3}$. Our ammonia Boltzmann plots also
suggest that there are several components with different
temperatures within our beams. This is also what we observe toward
the central region of our Galaxy (H\"uttemeister et al.
\cite{hue93b}).
\begin{table}
\caption{Masses and NH$_3$ abundances of the warm ($\ge 150\,\rm
K$) gas component} \label{hotgas}
\begin{tabular}{l l l l l l}
\hline & $M_{\rm warm}^a$&$M_{\rm total}^b$ &$\frac{M_{\rm
warm}}{M_{\rm total}}$&log$X$(NH$_3$)$^c$\\
&\multicolumn{2}{c}{$10^6$\,M$_\odot$} & \\
\hline NGC\,253&2.5&75&0.03&$-6.1$\\
IC\,342& 1.3 & 13 & 0.10 & $-7.5$\\
M\,82 & 75$^d$ & 100 & 0.2 \ldots 0.8$^d$ & $< -9$\\
GC$^e$& & & 0.3& $\ga -7$\\\hline
\end{tabular}\\
\begin{footnotesize}a) Within a $14''\times 27''$ aperture, from Rigopoulou et al
(\cite{rigopoulou02}) scaled to the distances in
Table\,\ref{summary}\\ b) from $^{13}$CO data in this paper, the uncertainty is a factor of 2\\
c) The warm NH$_3$ relative to the warm H$_2$ \\d) It is unknown
to us to which pointing position and orientation of the ISO beam
in Rigopoulou et al. (\cite{rigopoulou02}) the mass in Col.\,2
refers;
therefore the large uncertainty of Col.\,4\\
e) Rodr\'{\i}guez-Fern\'andez et al. (\cite{rodriguez01})\\
\end{footnotesize}
\end{table}
The relative abundances of the high density tracers NH$_3$ and CS
in Table\,\ref{summary} are from a comparison with the total
molecular column densities derived from $^{13}$CO; they are
therefore an average over all gas components. We can, however,
estimate the relative abundance of warm ammonia in the gas
component where it probably arises. This component is traced by
rotational quadrupole transitions of H$_2$, e.g. the $0-0$ $S(0)$
line, the upper level of which is 510\,K above the ground state.
From such lines masses of warm gas with a temperature of $\sim
150\,\rm K$  have been recently determined by Rigopoulou et al.
(\cite{rigopoulou02}) for a number of starburst and Seyfert
galaxies.  We have scaled the masses quoted in Rigopoulou et al.
(\cite{rigopoulou02}) to the distances used here and list them in
Table \ref{hotgas}. When we compare the masses of warm H$_2$
toward M\,82, IC\,342, NGC\,253 with the total H$_2$ masses from
our $^{13}$CO data and also take into account Galactic Center
H$_2$ measurements obtained by Rodr\'{\i}guez-Fern\'andez et al.
(\cite{rodriguez01}) it turns out that the warm gas makes up a
substantial fraction of the total gas mass in the centers of
nearby starburst galaxies and our own Galaxy. We cannot confirm,
however, the claim by Bradford et al. (\cite{bradford02}) that
\textit{all} the gas traced by CO is warm (and, hence, requires a
uniformly heating source such as cosmic rays) since this would
lead to contradicting H$_2$ mass estimates from observations of
H$_2$ and CO.

Newly formed high mass stars may heat only very localized regions
via dust absorption. Even in the central regions of luminous
galaxies, global dust temperatures rarely exceed 50\,K (e.g.
Henkel et al. \cite{Henkel86} ). In the case of NGC\,253,  the
central mid-infrared emission comes mainly from cold (23\,K) dust;
only a small fraction, $\ll20\%$ of the dust mass has temperatures
similar to 148\,K (Melo et al. \cite{melo2002}). Possible
mechanisms capable of elevating the H$_2$ temperature to global
values of 100\,K and more over an extent of several 100\,pc are
shocks, which may result e.g. from the dissipation of tidal
motions, photodissociation  and irradiation by X-rays (see e.g.
Rigopoulou et al. \cite{rigopoulou02}). Other possible heating
mechanisms involve cosmic rays (G\"usten et al. \cite{guesten81},
Farquhar et al. \cite{farquhar94}) or a magnetic ion-slip process
(Scalo \cite{scalo77}).

In the central regions of IC\,342 and NGC\,253, the relative
abundance of the hot ammonia to the hot H$_2$ component is
elevated over the relative ammonia abundances observed also
including the cold component. In NGC\,253 a value of $10^{-6}$ is
reached, which is similar to values found in Galactic hot cores
(e.g. H\"uttemeister et al. \cite{hue93b}) and the central region
of the Milky Way (Rodr\'{\i}guez-Fern\'andez et al.
\cite{rodriguez01}). The high NH$_3$ abundances in these galaxies
and presumably also in Maffei\,2 exclude heating mechanisms, such
as photo-ejection from grains (Genzel et al. \cite{genzel84}),
capable of destroying this fragile molecule which has a
dissociation energy of only 4\,eV. NH$_3$ has been observed, with
a similar abundance as in the warm components of NGC\,253 and
IC\,342, in Wolf Rayet environments (Rizzo et al. \cite{rizzo01}),
which can be explained by a release of NH$_3$ from dust grains
facilitated by weak shocks from the expansion of the bubble
created by the stellar wind. C-shocks have also to be used to
explain the abundance of complex molecules and the heating in the
central region of our Galaxy (Flower et al. \cite{flower1995},
Mart{\'\i}n-Pintado et al. \cite{martin2001}). Such intermittent
shocks, created by frequent cloud-cloud collisions can also
explain the temperature gradients seen in the NH$_3$ data. Cosmic
rays can also heat large volumes of gas to high temperatures

It is interesting that the warm gas fraction seems to be highest
in M\,82 where no warm ammonia component has been detected. Here
the relative ammonia abundance in the hot gas component must be $<
10^{-9}$.  This indicates that NH$_3$ traces a different molecular
gas component in M\,82 than in NGC\,253, Maffei\,2 and IC\,342.
For M\,82, the presence of an intense dissociating radiation field
and the predominance of warm molecular filaments with low H$_2$
column density and density (Mao et al. \cite{mao2000}) do not only
explain the low abundance of NH$_3$ (see e.g. the model in Fuente
et al. \cite{Fuente93}) and other complex molecules but also the
excitation of CO rotational levels (Wei\ss{} \cite{weiss01b}) in
the warm  H$_2$ gas detected by Rigopoulou et al.
(\cite{rigopoulou02}). One could argue that larger quantities of
warm NH$_3$ exist in this low density gas, but are escaping from
being detected due to lacking excitation. This is, however,
improbable since NH$_3$ molecules with their low dissociation
energy would be rapidly destroyed by the interstellar radiation
field in a low H$_2$ density and column density environment.

The low NH$_3$ abundance contrasts with a CS abundance which is
normal or even slightly higher than in other galaxies. Chemical
models of photon dominated regions (PDR) show that at moderate
extinctions of 6---8 mags., CS is strongly enhanced up to values
of more than 10$^{-6}$ due to the presence of a small amount of
ionized sulphur (Sternberg \& Dalgarno \cite{sternberg95}). The
bulk of the CS we are seeing in M\,82 might come from such
moderately shielded regions.

The lack of ammonia in the warm phase of M\,82 also suggests that
the overall heating mechanism for the molecular gas is different.
In a starburst, photo-dissociating radiation is capable of heating
a large fraction of the gas to 150\,K (e.g.
Rodr\'{\i}guez-Fern\'andez et al. \cite{rodriguez01}).  The low
abundance of ammonia toward M\,82, the properties of the CO
emission (Mao et al. \cite{mao2000}) and the detection of abundant
emission of HCO (Garc\'{\i}a-Burillo et al.
\cite{garciaburillo2002}), a tracer of photo-dissociation regions,
indicates that  PDRs are the main heating source that also have
influenced the chemistry of the bulk of the circumnuclear gas in
this galaxy. In regions where the CS abundance is enhanced,
shielding may be sufficient to prevent the heating of the CS
emitting gas to high kinetic temperatures.

\section{Conclusions}
The main conclusions of this paper are:

1. Our detections of highly excited lines of ammonia toward the
nuclear regions of three nearby galaxies reveal the presence of
gas components as warm as 50\,K \ldots 440\,K toward Maffei\,2,
IC\,342 and NGC\,253. Part of the gas may be cooler ($\sim
50\,$K). These temperatures resemble values observed toward the
central region of our own Galaxy.

2. The global beam averaged relative ammonia abundances toward
NGC\,253, Maffei\,2 and IC\,342 are with (3.3 \ldots 5.7)
$10^{-8}$ remarkably similar. The NH$_3$ abundances in the warm
($\ga 150\,\rm K$) phase in those galaxies, and also in the
central region of the Milky Way, are $\sim 10^{-7}$. Toward M\,82
this abundance is lower by more than an order of magnitude.

3. Toward NGC\,253, IC\,342 and M\,82 an elevated H$_2$ density of
$\sim 10^{4}$\,cm$^{-3}$ is estimated from multilevel studies of
CS.

4. The rich chemistry and high ammonia abundances observed in
NGC\,253, Maffei\,2, IC\,342 and the Galactic central region may
be in part explained by evaporation from the ice mantles of
interstellar dust grains due to C-shocks. These shocks would also
provide the heating to the observed elevated temperatures.

5. In the central region of M\,82, photodissociating radiation can
explain the high temperature of H$_2$, the lack of ammonia and
other complex molecules in M\,82, and the high abundance of CS and
HCO.

\begin{acknowledgements}
We thank Drs. David Meier and Jean Turner for making available
their $^{13}$CO $J=1-0$ data to us. Dr. R. Peng made available in
digital form their CS data toward NGC\,253. We thank the staff at
the Effelsberg 100-m telescope and the Heinrich-Hertz 10-m sub-mm
telescope for their valuable support during the observations and
wish to thank an anonymous referee for constructive criticism that
helped to improve the article.
\end{acknowledgements}

\end{document}